\documentclass[letterpaper]{jpconf}
\usepackage{graphicx}
\usepackage{subfigure}
\begin{document}
\title{Searching for physics beyond the Standard Model through the dipole
  interaction}
\author{B. Lee Roberts}
\address{Department of Physics, Boston University, Boston, MA 02215, USA}

\ead{roberts@bu.edu}

\begin{abstract}
The magnetic dipole interaction played a central role in the development of
QED, and continued in that role for the Standard Model. The muon
anomalous magnetic moment has served as a benchmark for models of new
physics, and the present experimental value is larger than the standard-model
value by more than three standard deviations. The electric
dipole moment (EDM) violates parity ({$P$}) and time-reversal ({$T$}) 
symmetries, and
in the context of the $CPT$ theorem,
 the combination of charge conjugation and parity ($CP$).  Since a
new source of {$ CP$} violation outside of that observed in the $K$ and 
$B$ meson systems is needed to help explain the baryon
asymmetry of the universe,  searches for EDMs are
 being carried out worldwide on a
number of systems.    The standard-model value of the EDM is
immeasurably small, so any evidence for an EDM would signify the observation
of new physics.  Unique opportunities exist for
 EDM searches using
polarized proton, deuteron or muon beams in storage rings. 
 This talk will provide an overview of the theory of dipole moments, and
the relevant experiments. The connection to 
the transition dipole moment that could produce lepton flavor violating 
interactions such as $\mu^+ \rightarrow e^+ \gamma$ is also mentioned.
\end{abstract}

 
\section{Introduction}

Measurements of dipole moments
 have played an important role in our understanding of
the subatomic world. Contrary what we teach our undergraduates in
modern physics, the Stern-Gerlach experiment did not
motivate the invention of spin, however
 the modern interpretation of their result is 
that the magnetic moment of the electron is one
Bohr-magneton~\cite{blr-robertsLM} with a $g$-value of two. 
While ``a spinning electron'' was proposed by Compton
 to explain ferromagnetism~\cite{blr-Compton21},
it was the introduction of spin and the associated magnetic moment
by Uhlenbeck and 
Goudsmit
to explain the fine-structure in atomic spectra~\cite{blr-Uhlenbeck25-26}
 that was
the beginning of spin physics as we now know it~\cite{blr-robertsLM}.

The discovery of anomalous magnetic moments was a critical event for 
twentieth-century physics that began with the  observations
that the hyperfine structure of hydrogen (HHFS) was too
large~\cite{blr-HHFS} when compared to the standard (Dirac) 
theory~\cite{blr-Dirac28}.
The Dirac equation\footnote{Throughout this paper I adopt the convention
 $e>0$.} 
\begin{equation}
i\left(\partial_{\mu}-ieA_{\mu}\left(x\right)\right)\gamma^{\mu}\psi\left(x\right)
=m_{e}\psi\left(x\right), 
\label{eq:dirac}
\end{equation}
predicted a magnetic dipole moment (MDM) for the electron 
\begin{equation}
\vec{\mu}_{e}=g_{e}\left( \frac{Qe}{2m_{e}}\right) \vec{s}\, ,
\ {\rm where\ the \ anomaly, \ defined\ as\ } \  a_e = \frac {(g_e-2)}{2}\, , 
\ {\rm is\ zero}
\label{eq:1.2}
\end{equation}
since in Dirac theory $g_e \equiv 2$. The increase in the 
hydrogen hyperfine levels could be interpreted as coming from an
additional magnetic moment. 
Motivated by the HHFS dilemma,  Schwinger~\cite{blr-schwinger}
 carried out what today is called ``the first loop
calculation'', and predicted that the electron had an additional
(anomalous) magnetic moment, $a_e = {\alpha}/{2\pi}$.
  The concurrent
precision spectroscopy measurements of Kusch and
 Foley~\cite{blr-Kusch}, obtained a
value for $g_e$ that was in good agreement with Schwinger's prediction. 

It took some years before the equivalent measurement was made for the muon.
The spin-rotation experiment of Garwin, et al.~\cite{Garwin-57},
which was one of the pioneering experiments~\cite{Garwin-57,Friedman-57} 
that observed parity violation in muon 
decay, found that the observed 
rate of spin rotation gave  $g_\mu =2.0 \pm 0.10$.  This result indicated 
``the very strong probability that the spin of the $\mu^+$
is $\frac{1}{2}$,'' thereby providing 
the first  indication that the muon behaved like a heavy 
electron.
A second muon spin rotation experiment by Garwin {\it et al.}~\cite{Garwin-60},
obtained a 12\% measurement of
the muon anomaly, $a_\mu^+ = 0.001\,13_{-0.000\,12}^{+0.000\,16}$ which agreed 
very well with the expected Schwinger value of 
$\alpha/2\pi \simeq 0.001\,161\dots$.  This
experiment showed conclusively that the muon did indeed have the
characteristics of a heavy electron.   

In 1950, well in advance of the famous Lee-Yang paper~\cite{Lee-56mu},
 Purcell and Ramsey~\cite{blr-Purcell} observed that there was no evidence for 
parity conservation in the nuclear force and that
 an electric dipole moment (EDM) of the neutron would
violate parity invariance {\sl P}. 
 This was of course the correct New-Physics effect  to
 look for, but in the wrong
place. Their initial experiment~\cite{blr-smith57}, which went unpublished
until after the discovery of parity violation, achieved a limit of 
$\vert d_n \vert < 5 \times 10^{-20}\, e\cdot$cm, a 
null result which has been pushed 
down to $2.9 \times 10^{-26}\, e\cdot$cm during the subsequent fifty-some years.
It was realized in 1957~\cite{blr-landau57,blr-ramsey58}
 that an EDM would also violate time-reversal
symmetry, {\sl T}, and by implication {\sl CP}.  This can be seen by
considering the Hamiltonian for dipole interactions:
$H = - \vec \mu \cdot \vec B - \vec d \cdot \vec E$.
The magnetic moment transforms like a spin (a pseudovector) as does the
EDM. The electric field is a vector while  the magnetic field is a
pseudovector.  
The combination $\vec \mu \cdot \vec B$ is even under {\sl C, P} and 
{\sl T}.  The EDM term,  $\vec d \cdot \vec E$, is odd under {\sl P} and {\sl
  T}, thus they are not good symmetries of this Hamiltonian.
  Presumably, new, as yet
undiscovered sources of {\sl CP} violation are responsible for the
matter-antimatter asymmetry in the universe, and would partially explain why
we are here.

\section{The Dipole Operators}

As mentioned above, the Dirac equation is inadequate to describe the measured
magnetic moment of the electron.  It is necessary to add a ``Pauli'' term
\begin{equation}
 \frac{Qe}{4m_{e}}a_{e}F_{\mu\nu}(x)\sigma^{\mu\nu}\psi(x)
\label{eq:blr-pauli}
\end{equation} 
which in modern language is a dimension 5 operator that must arise from
loops in a renormalizable theory.  New Physics (NP) can also contribute
through loops, with $a({\rm NP})= C(m/\Lambda)^2$ where 
$C \simeq {\mathcal O}(1)$, or $\simeq {\mathcal O}(\alpha)$ in
weak coupling loop scenarios.
In the same spirit, one could add the following Pauli-like term
\begin{equation}
\frac{i}{ 2}
d_{e}F_{\mu\nu}\left(x\right)\sigma^{\mu\nu}\gamma_{5}\psi\left(x\right)
\ \ {\rm with} \ \ \vec d = \eta \left( \frac{Qe}{2mc}\right) \vec s,
\label{eq:EDM}
\end{equation}
which represents the electric dipole moment interaction.
The quantity $\eta$ plays the role for the EDM that $g$ plays for the
MDM.

The electromagnetic current is 
$\left\langle
  f\left(p'\right)\left|J_{\mu}^{em}\right|f\left(p\right)\right\rangle
=\bar{u}_{f}\left(p'\right)\Gamma_{\mu}u_{f}\left(p\right)$
where $\bar{u}_{f}$ and $u_{f}$ are Dirac spinor fields and $\Gamma_{\mu}$ has
the general Lorentz structure
\begin{equation}
\Gamma_{\mu} = 
F_{1}\left(q^{2}\right)\gamma_{\mu}
+iF_{2}\left(q^{2}\right)\sigma_{\mu\nu}q^{\nu}
-F_{3}\left(q^{2}\right)\sigma_{\mu\nu}q^{\nu}\gamma_{5} 
+F_{A}\left(q^{2}\right)\left(\gamma_{\mu}q^{2}-2m_{f}q_{\mu}\right)\gamma_{5}; 
\label{eq:2.2}
\end{equation}
with $F_1(0) = Qe$ the electric charge, $F_2(0) = a(Qe/2m)$ the anomalous 
magnetic moment (anomaly), and $F_3 = dQ$ the electric dipole moment. 
 I  ignore the $F_A$ term, the anapole moment.

The anomalous part of the dipole moment interaction
\begin{equation}
\bar u_{\mu}\left[Qe F_1(q^2)\gamma_{\beta} +
{i Qe \over 2m_{\mu}}F_2(q^2)\sigma_{\beta \delta}q^{\delta}\right]u_{\mu}
\label{eq:blr-dipole}
\end{equation}
connects states of opposite helicity, i.e. it is chiral changing, giving
 it a unique sensitivity to New Physics interactions, e.g. the sensitivity to 
$\tan \beta$ in supersymmetric (SUSY) theories. In most SUSY models,
the contribution to $a_\mu$ depends on the SUSY mass scale, the sign of the 
$\mu$ parameter, and $\tan \beta$. A simple SUSY model
with equal masses~\cite{blr-CzarLM,blr-Czar01} gives the SUSY contribution as:
\begin{equation}
\simeq ({\rm sgn} \mu) \ 130 \times 10^{-11}\ \tan \beta\ 
\left(\frac{100\ {\rm GeV}  }{ \tilde m}\right)^2
\label{eq:blr-susy}
\end{equation}

As mentioned above, an EDM violates both {\sl P} and {\sl T} symmetries
and by implication {\sl CP}. 
For hadronic systems, the ``theta'' term in the QCD Lagrangian\footnote{ In
  electromagnetic theory, the equivalent sort of term is $\vec E \cdot \vec
  B$, which also is odd under {\sl P} and {\sl T}. Such a term is important in
describing topological insulators in condensed matter physics.}
\begin{equation}
{\mathcal L }^{eff}_{QCD} = {\mathcal L}_{QCD} + \theta \frac{g^2_{ QCD}}{32 \pi^2}F^{a \mu\nu}
\tilde F_{a \mu \nu }  \ \ a = 1,2,\dots,8
\label{eq:blr-theta}
\end{equation}
violates both parity and time-reversal symmetries,
where the physical quantity is the sum of $\theta$ and the overall phase in
the quark matrix, 
 $\bar \theta = \theta + \arg(\det M)$.
The non-observation of a neutron EDM restricts the value of $\bar \theta$:\ \ 
$\left|d_{n}\right|\simeq3.6\times10^{-16} \bar{\theta}\,
e\cdot\mathrm{cm}\ \Rightarrow \ \bar{\theta} \leq  10^{-10}\,$,
which for a quantity that could be order one is anomalously small. The
smallness of $\bar \theta$ 
is often referred to as {\it the strong {\sl CP} problem}.  
While supersymmetry, 
or other models of New Physics can 
easily contain new sources of {\sl CP} violation, the absence of any observation
 of an EDM, with a significant
 fraction of the ``natural'' part of the SUSY {\sl CP}-violating 
parameter space already eliminated, is sometimes called
{ \it  the SUSY {\sl CP} problem}.

The isovector and isoscalar combinations of the magnetic dipole
moments are:
\begin{equation}
F_{2N}^{(I=1)} = \frac{F_{2p} - F_{2n}}{2} \simeq 1.85 \,; \qquad 
F_{2N}^{(I=0)}  = \frac{F_{2p} + F_{2n}}{2} \simeq -0.06 \, ;
\end{equation}
we conclude that the isovector dominates the anomalous MDM.  Both isoscalar
and isovector EDMs are predicted by various models~\cite{blr-posp-ritz}, 
so measuring both the
proton and neutron EDMs would help disentangle these two possibilities.

\section{Measurements of Dipole Moments}

\subsection{Measurements of the Muon and Electron Anomalies}

The electron anomaly has been measured to a precision of 0.24 parts
billion by storing a single electron in a quantum cyclotron
and measuring the quantum cyclotron and spin levels 
in this system~\cite{blr-hanneke08}.
Were an independent measurement of the
fine-structure constant $\alpha$ available at this precision, this
impressively precise measurement could provide a testing ground for the
validity of QED down to the five-loop level, and present an opportunity to
search for effects of New Physics. At present the best independent
measurements of $\alpha$ have a precision of 
$\sim 5$~ppb~\cite{blr-alpha,blr-Gabrielse10}.
  In the absence of such an independent
measurement, the electron $(g-2)$ value has been used, along with the QED
theory (assumed to be valid), to give the most precise 
value of $\alpha$~\cite{blr-hanneke08}. 

The muon anomaly, which has been measured to an accuracy of 0.54 parts per
million (ppm)~\cite{blr-bennett04},  has an increased 
sensitivity to heavier physics 
that scales as $(m_\mu/m_e)^2 \simeq 43,000$. This means that at a
measurable level the Standard-Model contributions to the muon anomaly come
from QED; from virtual hadrons in vacuum polarization or hadronic light-by-light
scattering loops; and from loops involving the electroweak gauge bosons.

In principle the technique is similar to the measurement of the electron
anomaly,  where muons are stored in a ``trap''
consisting of a dipole magnetic field plus an electrostatic quadrupole
field. In the muon experiment, an ensemble of muons is injected into
a precision storage ring.  The observable is the spin precession
frequency relative to the momentum, which is the difference between the spin
precession frequency and the cyclotron frequency:
\begin{equation}
\vec \omega_a = \vec \omega_S - \vec \omega_C 
= - \ \frac{Qe }{ m} 
\left[ a_{\mu} \vec B -
\left( a_{\mu}- \frac{1 }{ \gamma^2 - 1} \right)
\frac{ {\vec \beta \times \vec E }}{ c }
\right]\,.
\label{eq:blr-omega}
\end{equation}
The second term in brackets represents the effect of the motional magnetic
field on the spin motion. The experiment is operated at the ``magic'' value
of $\gamma_{magic} = 29.3$ where this motional term vanishes, which permits
the use of an electric quadrupole field to provide the vertical focusing. 

\begin{figure}[h]
\centering
\subfigure[]{\includegraphics[width=0.3\textwidth]{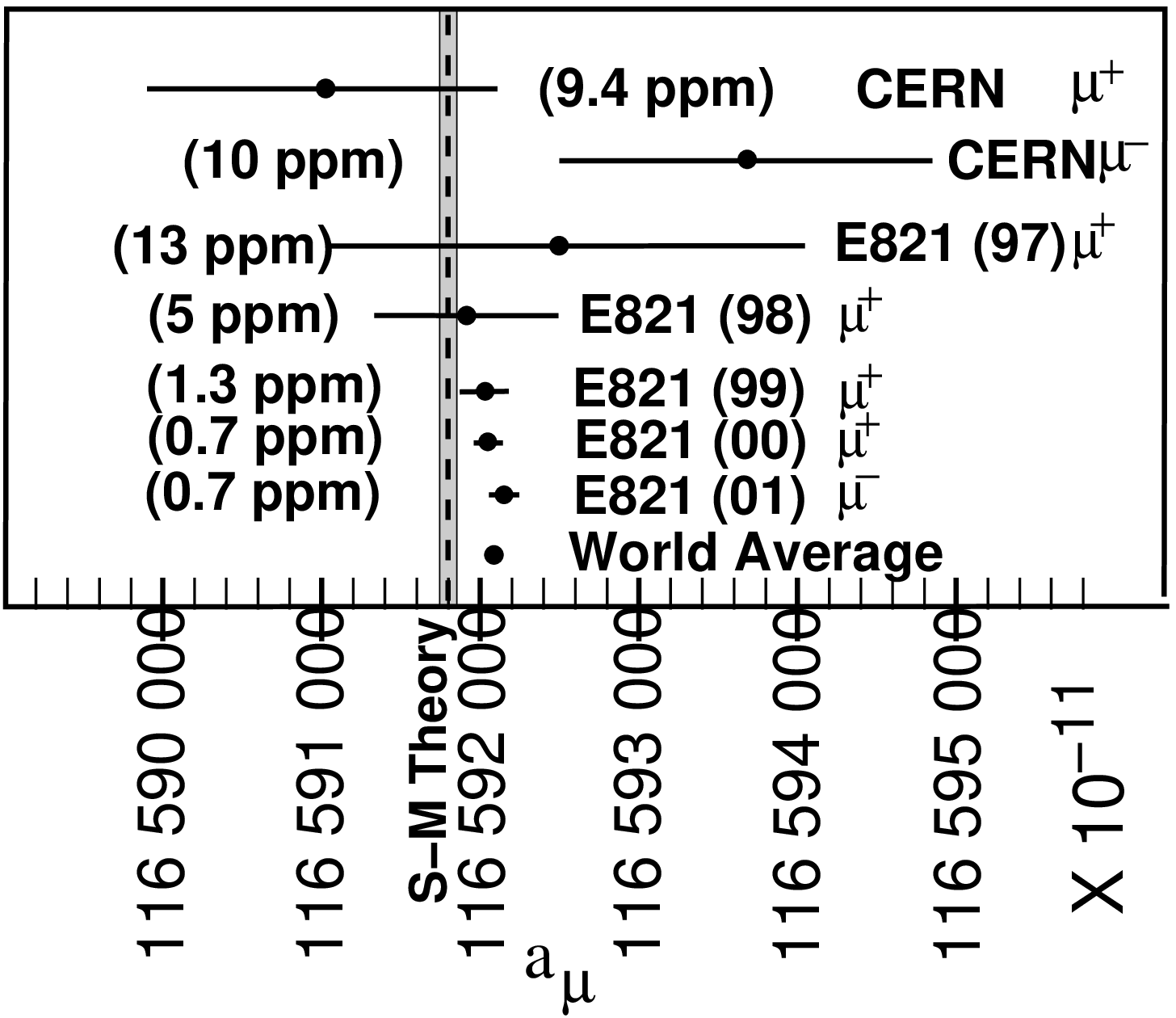}}
\subfigure[]{\includegraphics[width=0.3\textwidth]{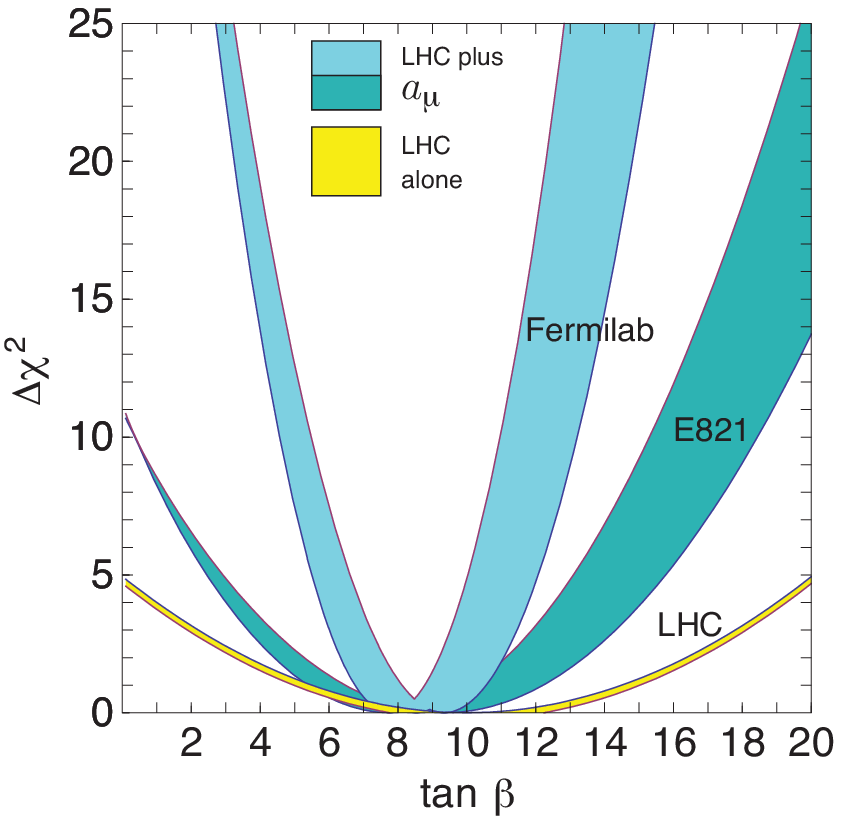}}
\caption{(a) Measurements of the muon anomalous magnetic moment. 
The theory value
shown is taken from Ref.~\cite{blr-Davier-tau10} as described in the text.
(b) Implications for a $\tan \beta$ determination assuming the LHC were
  to discover the SPS1a SUSY scenario,
which predicts $\Delta a_\mu = 293 \times 10^{-11}$.  The 
$a_\mu$ curves assumed
 $\Delta a_\mu = (255 \pm 80) \times 10^{-11}$,
the difference before new data were included in
the evaluation of the hadronic contribution~\cite{blr-Davier-tau10}.
(Figure courtesy of Domink St\"ockinger)}
\label{fg:blr-amu}
\end{figure}

The measured electron and muon anomalies are
\begin{eqnarray}
a_e &=& [115\, 965\, 218\, 073 (28)] \times 10^{-14} \,0.24\,{\rm ppb}\\
a_\mu &=& [116\,592\,089 (63)] \times 10^{-11}\, 0.54\,{\rm
  ppm} \label{eq:amuexp} 
\end{eqnarray}
The individual measurements that go into $a_\mu$ are shown in
Fig.~\ref{fg:blr-amu}(a).

The comparison between the Standard-Model and experimental values of $a_\mu$
 is partially limited by the knowledge of the 
 hadronic contribution.
Significant work on different aspects of the
hadronic contribution are in progress, both on the experimental side to
measure the hadronic electroproduction cross sections better, and on theoretical
efforts to improve on the hadronic light-by-light
contribution~\cite{blr-phipsi09}.
There appears to be a difference 
$\Delta_{a_\mu} = (287 \pm 80)\times 10^{-11}$,  $3.6~\sigma$, between
the experimental value (Eq.~\ref{eq:amuexp}) and the Standard-Model value
of~\cite{blr-Davier-tau10,blr-PdRV}
$a_\mu^{\rm SM}[e^+e^-] = 116\, 591\, 802 (49) ]\times 10^{-11}$.

Such a deviation could
 fit well with the expectations of supersymmetry in the few-hundred GeV mass
 region,  as shown in
Eq.~\ref{eq:blr-susy}.
Were SUSY particles to be discovered 
at LHC, the muon anomaly would play an important role in helping to
discriminate between the different possible scenarios, and providing
a measure of $\tan \beta$. For a thorough 
review of SUSY and $(g-2)$ see the 
articles by St\"ockinger~\cite{blr-stockinger}.

The precision of the E821 $(g-2)$ measurement was limited by the statistical
error of 0.46~ppm, compared to the systematic error of 0.28~ppm.  A new
experiment has been proposed for Fermilab, P989~\cite{blr-p989} with the goal of
equal statistical and systematic errors, and a total error of 0.14 ppm, a
factor of four improvement over E821. 

The supersymmetry community has chosen a number of possible scenarios 
that might be discovered at LHC, the Snowmass points and slopes~\cite{blr-SPS},
which serve as benchmarks for determining the sensitivity to the SUSY
parameters.
Since $a_\mu$ has significant sensitivity to $\tan \beta$ 
(see Eq.~\ref{eq:blr-susy}), it is possible to compare the sensitivity to
$\tan \beta$ from LHC vs. from $\Delta a_\mu$. Such a comparison is shown in
Fig.~\ref{fg:blr-amu}(b), which
 assumes that the SPS1a point is realized, 
a typical mSUGRA point with an intermediate value of $\tan \beta$. 
   The lighter blue band shows the improvement
that could be gained in the new Fermilab experiment.

\subsection{Electric Dipole Moments}

Unlike the magnetic dipole moments, the Standard-Model values of electric
dipole moments are
 orders of magnitude less than present experimental limits, both of which
are shown in Table~\ref{tb:blr-edm}. The the
experimental observation of an EDM would unambiguously 
signify the presence of new
physics.

\begin{table}[h!]
\caption{Measured limits on electric dipole moments, and their Standard
Model values}
\label{tb:blr-edm}
\begin{tabular}{|c|c|c|} \hline
   { Particle}  &{ Present EDM} & { Standard Model}  \\
                 & Limit ($e\cdot$~cm)           & Value ($e\cdot$~cm) \\
\hline
$n$ & {$2.9 \times 10^{-26}$ } (90\%CL)~\cite{blr-nEDM}  & {$\simeq 10^{-32}$ }  \\
\hline
$p$  & $ 7.9 \times 10^{-25}$~\cite{blr-hgEDM} &  {$\simeq 10^{-32}$ }  \\
\hline
 $e^-$  & {$\sim 1.6 \times 10^{-27 }$} (90\%CL)~\cite{blr-eEDM} & {$10^{-38}$ } \\
\hline
{$\mu$} &{$1.8 < 10^{-19}$ } (95\%CL)~\cite{blr-bennett09} & {$10^{-35}$ }\\
\hline
$^{199}{\mathrm Hg}$ & $ 3.1 \times  10^{-29}$  (95\%CL)~\cite{blr-hgEDM} & \\
\hline
\end{tabular}
\end{table}

\begin{figure}
\centering
\subfigure[]{\includegraphics[width=0.6\textwidth]{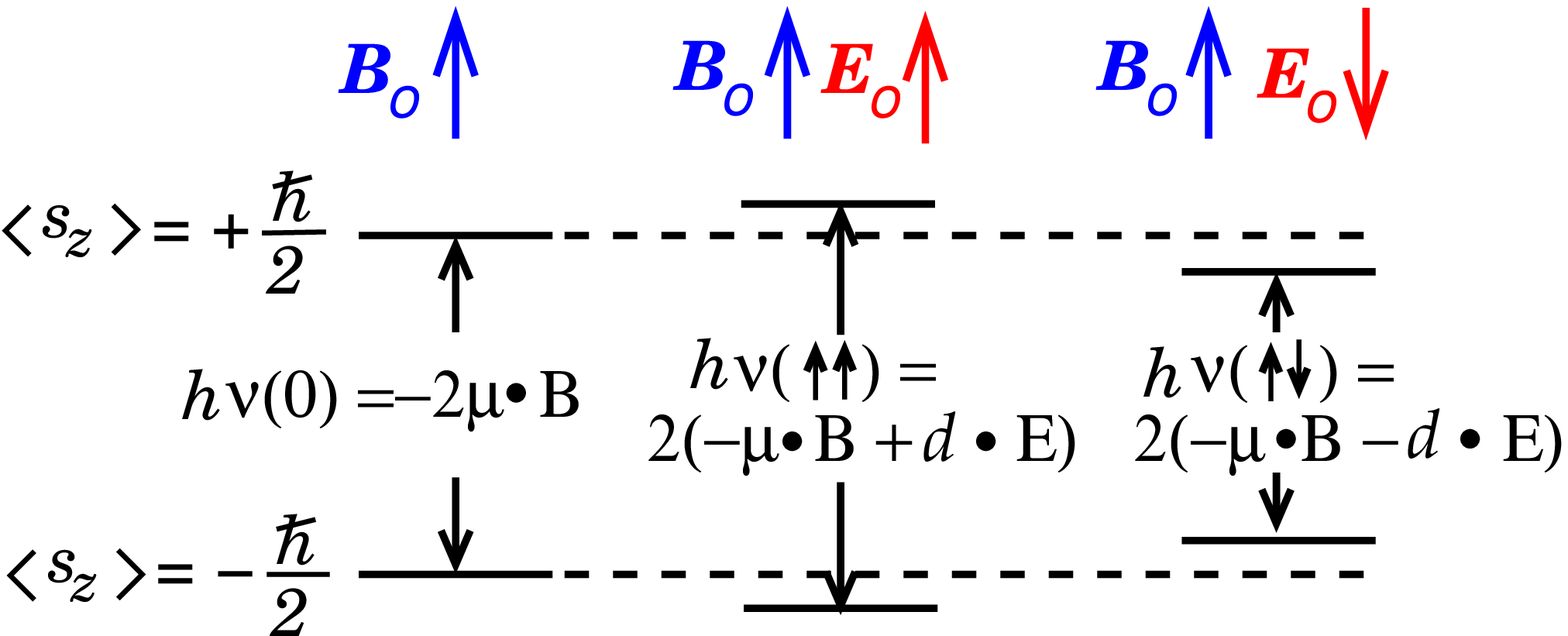}}
\subfigure[]{\includegraphics[width=0.3\textwidth]{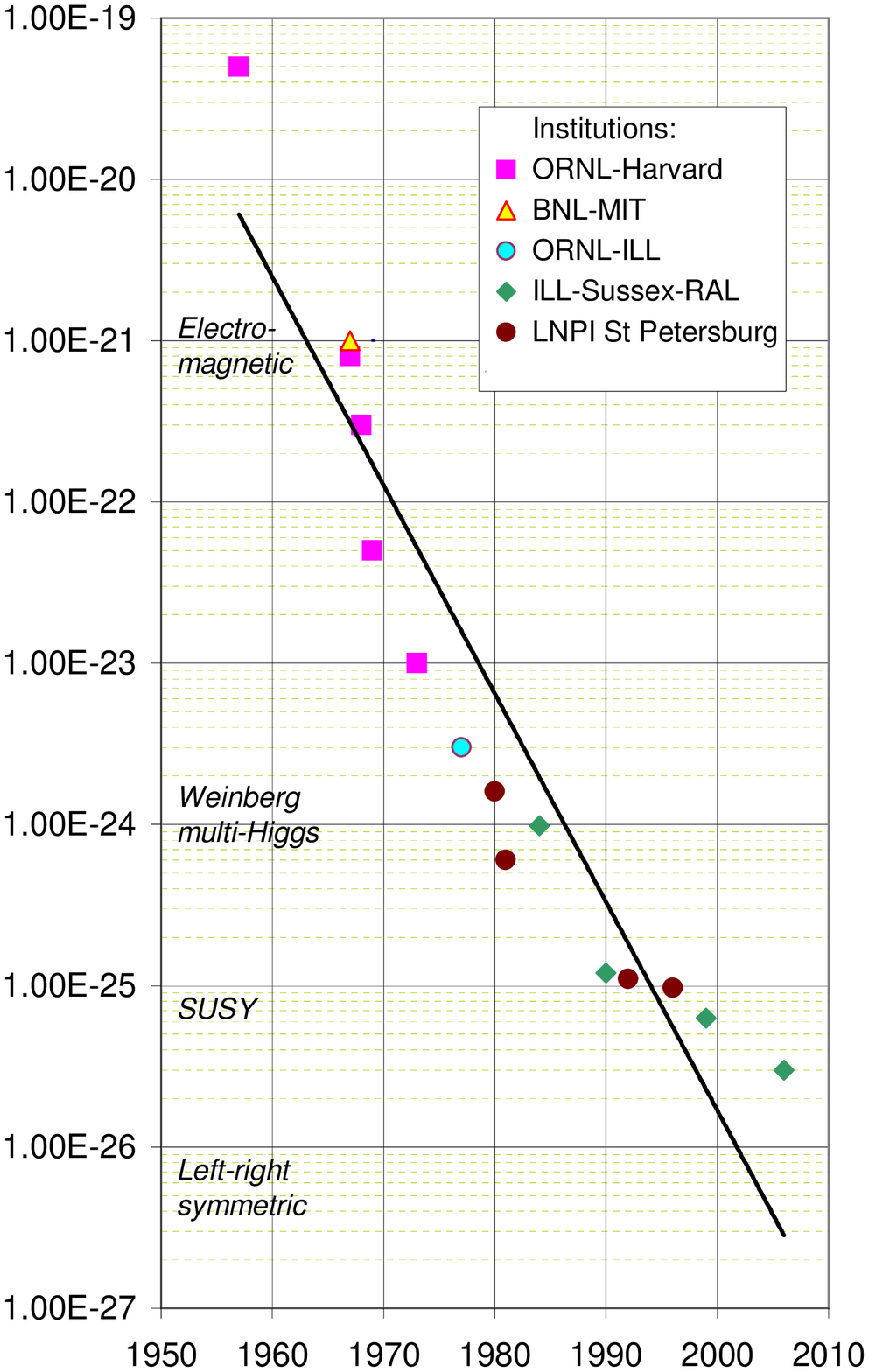}}
\caption{(a) The principle of EDM experiments.
(b) Limits on the neutron EDM as a function of time. (courtesy of
P.G. Harris).
\label{fg:blr-edms}}

\end{figure}

In the traditional EDM experiment, the system is placed in a region of 
parallel (anti-parallel) electric and magnetic fields
 (see Fig.~\ref{fg:blr-edms}(a)). The Larmor frequency is
measured, and then the electric field direction is flipped.  An EDM would
cause the Larmor frequency to be higher/lower depending on the direction of
the electric field.  The EDM is determined by the frequency difference
between these two configurations:
$\Delta \nu =\nu_{\uparrow \uparrow}- \nu_{\uparrow \downarrow} = ({4 d E})/({h})$.

The limit on the neutron EDM versus time is shown in
Fig.~\ref{fg:blr-edms}(b).
The lowest limit for the EDM of any system comes from the 
$^{199} {\rm  Hg}$ atom~\cite{blr-hgEDM}: 
$d(^{199} {\mathrm Hg}) = (0.49 \pm 1.29_{stat} \pm 0.76_{syst}) 
\times 10^{-29}\, e\cdot{\rm cm}$
giving the limit above in Table~\ref{tb:blr-edm}.

Searches are underway worldwide to find an EDM of the 
electron~\cite{blr-commins} (Imperial College, Colorado, Harvard, 
Yale, Amherst, Penn State,
Texas, Osaka and Indiana),
neutron~\cite{blr-lamoreaux} (ILL, PSI, Oak Ridge, TRIUMF), the 
atoms~\cite{blr-fortson} $^{199}$Hg (Seattle), 
$^{129}$Xe (Princeton, Michigan), $^{225}$Ra (Argonne, Groningen).  However,
one word of caution, only the neutron, proton, deuteron and muon EDMs can be
measured {\it directly}.  All of the other
 EDM measurements take place in atoms or molecules,
and thus their interpretation is subject to issues of atomic and molecular
physics. 

The limit on the muon EDM comes from E821 at Brookhaven~\cite{blr-bennett09}. If
an EDM exists, it is necessary to modify the spin precession formula of
Eq.~\ref{eq:blr-omega} with an extra term $\omega_\eta$,
\begin{equation}
\vec \omega_\eta =
 \eta \frac {Qe}{2m}
 \left[ \frac {\vec{E}} {c}  +  \vec{\beta} \times \vec{B} \right] 
\label{eq:blr-omegaeta}
\end{equation}
and the total spin precession frequency is $\vec \omega = \vec \omega_a +
\vec \omega_\eta$.  The motional electric field is proportional to 
$\vec \beta \times \vec B$, so
 the EDM results in an out-of-plane component of the
spin, where the (very small) tipping angle relative to $\vec \omega_a$ is
$\delta =\tan^{-1} {\omega_{\eta}}/ {\omega_a}
= \tan^{-1} ({ \eta \beta }/{ 2a}).$ 
The parameter $\eta$
is related to the EDM, $d$, by the relationship
\begin{equation}
d =  \left ( \frac {e\hbar} {4mc} \right) \eta\, \ {\rm for\  spin}\ \frac{1}{2}\, ;
\ {\rm  and }\ 
d =  \left ( \frac {e\hbar} {2mc} \right) \eta \, \ {\rm for\ spin}\ 1\ .
\end{equation}

  In the $(g-2)$ experiments, $\omega_\eta \ll \omega_a$ and the resulting
motion is an up-down oscillation with frequency $\omega_a$,
{\em out of phase} with the $(g-2)$
oscillation.  Such an experiment is largely limited by systematic
errors~\cite{blr-bennett09}, since the out-of-plane motion is masked by the 
large-amplitude spin precession from the magnetic moment.
Nevertheless, the new Fermilab $(g-2)$
 effort hopes to achieve one to two orders of
magnitude improvement in the muon EDM as a by-product of the improved $(g-2)$
measurement.  Significant progress beyond that goal 
would need to reduce the large
background caused by the $\omega_a$ precession.  

To achieve this reduction,
the ``frozen spin'' technique has been 
proposed~\cite{blr-farley-fs,blr-roberts-edm}.
 Recall the point of choosing the magic $\gamma$ in
Eq.~\ref{eq:blr-omega} was to eliminate the effect of the focusing electric
field on the spin precession.  If however, a storage ring were to be operated
at a different momentum, then a {\em radial} electric field could be 
used to counter the
 the spin precession from the magnetic moment (see Eq.~\ref{eq:blr-omega}),
 {\it viz.}
 it could be chosen such
 that $\omega_a = 0$. 
 The $E$-field required to freeze the muon spin is~\cite{blr-farley-fs} 
\begin{equation}
E=\frac {a B c \beta \gamma^2} {1 - a \beta^2 \gamma^2} \simeq aBc\beta
\gamma^2  .
  \label{eq:mrs-efield}
\end{equation}
 The frozen spin technique, along with a very
high-flux facility, could permit a sensitivity of $10^{-24}\, e \cdot$cm
 or better for the
muon EDM, providing a unique opportunity to measure the EDM of
 a second generation particle.

Both the proton EDM experiment
being proposed for Brookhaven, and the deuteron EDM experiment being
discussed for COSY present exciting and unique opportunities for direct
measurements of hadronic EDMs.  Perhaps it is obvious, but 
should convincing evidence for any EDM be found, it will be
imperative that as many other EDMs as possible
be measured to help sort out the
source of this new {\sl CP} violation.
More details on measuring EDMs in storage rings
 are given in the talk by Onderwater,
and in Refs.~\cite{blr-roberts-edm,blr-adelmann}.

\section{Transition Moments}

Although space limitations do not permit a detailed discussion of the
searches for lepton flavor violation, I do want to mention briefly the topic
of transition moments.
One of the most important discoveries in the past decade was the 
definitive evidence that neutrinos mix.  In the Standard Model, this implies
that charged leptons will also mix, however the calculated
transition rate for $\mu^+ \rightarrow e^+ \gamma $ is:
$Br(\mu \rightarrow e \gamma)=
({3 \alpha})/({32 \pi}) \left| \sum_\ell V^*_{\mu \ell} V_{e \ell}
 ({m_{\nu_\ell}^2})/({M_W^2})\right| ^2 
\leq 10^{-54}$,
which is immeasurable under the most optimistic experimental scenario. 
 Thus the observation of any process that violates lepton flavor,
such as $\mu^+ \rightarrow e^+ \gamma$, $\tau^+ \rightarrow \mu^+ \gamma$, or
coherent muon to electron conversion, 
$\mu^- + {\mathcal N} \rightarrow e^- +  {\mathcal N}$,
  would herald the discovery of new physics.

Just as the diagonal matrix elements of the electromagnetic current
were connected with the electric and magnetic dipole moments, we have
the off-diagonal elements of the current~\cite{blr-CzarLM} that give
transition moments:
$\left\langle
f_{j}\left(p'\right)\left|J_{\mu}^{\mathrm{em}}\right|
f_{i}\left(p\right)\right\rangle
 =  \bar{u}_{j}\left(p'\right)\Gamma_{\mu}^{ij}u_{i}\left(p\right),$
where $\Gamma_\mu^{ij}$ is given by
\begin{equation}
\Gamma_{\mu}^{ij}  = 
\left(q^{2}g_{\mu\nu}-q_{\mu}q_{\nu}\right)\gamma^{\nu}
\left[F_{E0}^{ij}\left(q^{2}\right)
+\gamma_{5}F_{M0}^{ij}\left(q^{2}\right)\right]
 +i\sigma_{\mu\nu}q^{\nu}\left[F_{M1}^{ij}\left(q^{2}\right)
+\gamma_{5}F_{E1}^{ij}\left(q^{2}\right)\right].
\label{eq:2.39}
\end{equation}
The first term gives rise to chiral-conserving flavor-changing amplitudes
at $q^2 \neq 0$, e.g. $K^+ \rightarrow \pi^+ e^+e^-$,
$\mu^+ \rightarrow e^+ e^+ e^-$, and the second (dipole) term gives rise to
chiral-changing, flavor-changing amplitudes, e.g. 
$b \rightarrow s \gamma$, $\mu \rightarrow e \gamma$ and $\tau \rightarrow
e \gamma$.  The search for these Standard-Model forbidden decays provides
a complementary path to discover new physics, and in some models the muon
anomaly, EDMs and charged lepton flavor violation are connected.
For more details, see the
reviews of charged lepton flavor 
violation~\cite{blr-marciano,blr-kuno,blr-kuno-okada,blr-okada}.

\section{Summary and Conclusions}

Spin physics began with the paper of Uhlenbeck and
Goudsmit~\cite{blr-Uhlenbeck25-26} which explicitly introduced the
concept of a magnetic moment associated with electron spin. 
Twenty five years later, Purcell and Ramsey~\cite{blr-Purcell}
 proposed to search for
an electric dipole moment to look for New Physics (parity violation).
We now recognize Dipole moments as an essential tool
in the search for physics beyond the
Standard Model.  

There may already be an indication of New Physics at the loop level from the
muon $(g-2)$ experiment. The new experiment to
 measure $a_\mu$ a factor of four more
precisely at Fermilab  should clarify the difference 
that has been observed between the
Standard-Model value and experiment.
 The observation of an electric dipole moment would
herald the discovery of a new source of {\sl CP} violation, which we believe
must exist to explain the matter-antimatter asymmetry of the universe. 
  Similarly, a discovery in
 the on-going searches  for
 charged lepton flavor violation at the Paul Scherrer Institut, Fermilab,
J-PARC, and the $B$ Factories would also herald New Physics at 
work in the charged lepton sector.

All of these experiments will help to guide our
interpretation of the new phenomena which we hope to discover at LHC.
There are a number of opportunities to make
important contributions to this field that are open to the spin
physics community. Of special note are the opportunities being
discussed for Brookhaven and for the COSY facility in J\"ulich 
to measure {\it directly} the
proton and deuteron EDMs using the storage-ring technique.

\subsection{Acknowledgments}

I wish to thank Dave Hertzog for comments on this manuscript, and to
acknowledge Andzrej Czarnecki, Michel Davier, David Hertzog,
Yoshi Kuno, Bill Marciano, Jim Miller, 
Eduardo de Rafael, Yannis Semertzidis and Domink St\"ockinger
for helpful conversations.  I thank Dominik St\"ockinger and Phil
Harris for the use of their figures.
The review article by Czarnecki and
Marciano in Lepton Dipole Moments~\cite{blr-CzarLM} was extremely 
helpful in preparing this talk.

\bigskip 

\end{document}